# Modeling of Growth Morphology of Core-Shell Nanoparticles

**Vyacheslav Gorshkov,**[a] **Vasily Kuzmenko,**[a] and **Vladimir Privman**[b,*]

[a] National Technical University of Ukraine — KPI, 37 Peremogy Avenue, Building 7, Kiev 03056, Ukraine

[b] Center for Advanced Materials Processing, Department of Physics, Clarkson University, Potsdam, NY 13699, USA

## Abstract

We model shell formation of core-shell noble metal nanoparticles. A recently developed kinetic Monte Carlo approach is utilized to reproduce growth morphologies realized in recent experiments on core-shell nanoparticle synthesis, which reported smooth epitaxially grown shells. Specifically, we identify growth regimes that yield such smooth shells, but also those that lead to the formation of shells made of small clusters. The developed modeling approach allows us to qualitatively study the effects of temperature and supply the shell-metal atoms on the resulting shell morphology, when grown on a pre-synthesized nanocrystal core.

---

[*]Corresponding author: e-mail privman@clarkson.edu; phone +1-315-268-3891

**I. Introduction**

We report results of a theoretical investigation that has been motivated by recent experimental observations[1,2] of the emergence of nanosize morphologies of noble metal shells which are grown on surfaces of gold nanoparticle cores. Generally, shells of core-shell nanoparticles as well as other surface structures can be grown by the processes of attachment and restructuring of deposits formed by atom-size matter: atoms, ions, molecules (to be collectively referred to as "atoms" for brevity), as well as by deposition of off-surface-nucleated clusters onto the surface. Here we explore the role of the former: atom-size matter kinetics, in the emergence of observed shell morphologies. Studies of synthesis of noble-metal and other core-shell particles[1-12] and generally nanostructure formation[13-23] in surface growth are active fields of experimental research motivated by numerous applications, e.g., in catalysis, fuel cells, and design of new materials. Recently published[1,2] and patented[24,25] experimental advances in noble-metal particle synthesis have yielded truly nanosize objects, with cores below 30–40 nm, and with smooth epitaxially grown shells as thin as 5–6 atomic layers, or, when not smooth, consisting of structures/clusters up to 3 nm. These findings pose interesting theoretical challenges. Here we utilize for core-shell geometries a modeling approach originally developed for nanoparticle shape selection,[26-28] and that was later applied to answer[29-30] qualitative and semi-quantitative questions for nanostructure growth problems on flat surfaces, such as which substrates and growth conditions are the best for growing specific morphologies, including the dependence of the growth process on the physical conditions: temperature, flux of matter, etc. This approach was also used[31] to study aspects of nanoparticle sintering.

Particles of sizes in the relevant ranges, of order a few tens on nanometers, contain too many atoms for a direct first-principle modeling of their cores' initial synthesis or their later overgrowth with shells. Furthermore, the dynamical processes involved: transport of atom-size matter to the surface, on-surface restructuring, and detachment/reattachment, are all ongoing in a surrounding medium, the effect of which should also be modeled. Therefore, the present problem in its entirety requires a multiscale approach. Indeed, theoretical modeling approaches are rather diverse[23,26-43] and are all focused on the features of the various growth problems at various scales, from few-atom clusters to statistical-mechanics models for large clusters, and to continuum descriptions for macroscopic scales.

For the scales relevant here, one important recent finding[26-28] has been that "persistency" can be a driving mechanism in the emergence of well-defined shapes and morphologies in nonequilibrium growth of nanoparticles and nanostructures. Here "nonequilibrium" refers to the overall rate of the local movements of atoms at the growing surface, means, that we are not in the regime when surface restructuring is fast enough (as compared to the rate of transport of new matter to the surface) to yield thermal equilibrium: This restructuring includes on-surface motion as well as detachment/(re)attachment, but *does not* refer to the global dynamics of matter transport in the system, such as the time-dependence of the flux of atoms to the surface. Initially, the property termed imperfect-oriented attachment[34,44-46] has been identified as persistency in successive nanocrystal binding events leading to the formation of uniform short chains of aggregated nanoparticles. Persistency can also mediate growth of other shapes and surface structures[26-30,46] from atoms, for a certain range of the resulting feature sizes. Nanosize structures simply do not contain enough constituent atoms for a large probability of fluctuations that result in internal defects or similar unstable surface features. The former are known to control the development of pronounced shape nonuniformities that would cause deviations from the so-called "isomeric," i.e., even-proportioned particle shapes, whereas the latter mediate the formation of fractal structures and/or "dendritic instabilities" of growing a hierarchy of side branches.[35,36]

Specifically, the kinetic Monte Carlo (MC) type model utilized here includes diffusional transport of atoms in space. They can attach to the growing particle shell, as well as move as part of the growing surface and detach/reattach from/to it, all this according to the thermal-like over-(free-)energy-barriers (Boltzmann factor) rules that will be described in the next section. These processes will determine the growing shell morphology. However, the on-surface and detachment/reattachment processes should not be fast enough to yield thermalization on the time scales of the transport of additional matter to the surface. Too fast, as compared to restructuring processes, a diffusional transport of matter to the surface would lead to fractal growth.[35,36] Too slow a transport would obviously result in the overall shape equilibration by restructuring processes, leading to thermal-equilibrium Wulff shapes.[37-40] Emergence of well-defined nanocrystal shapes for nanoparticles grown in the "nonequilibrium" regime when the time scales are comparable has recently been explored[26-28] in studies of isolated nanoparticle growth. Particle sintering,[31] as well as surface-feature formation,[29,30] have also been successfully studied within this approach.

The local "persistency" property is enforced in this model by requiring that, while diffusional transport occurs in the continuous three-dimensional space, atom attachment is only allowed "registered" with the underlying lattice structure, here taken as FCC for the relevant noble metals.[1,2] As explained in earlier works,[26,28-30] this rule mimics the property that growing structures are unlikely to develop "macroscopic" (particle-wide/structure-wide) defects as long as these structures are nanosize, which has been a property identified important for well-defined particle and feature shape selection in "nonequilibrium" growth at the nanoscale, with, for instance, the emerging particle shapes[26] defined by faces of the crystalline symmetry of the substrate, but with proportions different from those in the equilibrium Wulff growth.

The model is specified in Section II. Our goal has been to elucidate the extent to which this type of mesoscopic-scale kinetic modeling can qualitatively reproduce features observed in recent experiments,[1,2] including the possibility of finding kinetic conditions for growth of thin well-defined epitaxial shells, as well as of shells consisting of clusters, and certain other properties to be described in Section III, which is devoted to presentation and discussion of the results and to concluding remarks.

## II. Description of the Modeling Approach

The present model follows earlier works[26,29,30] in the setup of the kinetic MC approach, and therefore some of the model details will only be outlined here. Other model aspects, those specific to the considered core-shell geometry, will be described in greater details. We consider the role of the processes where by "atoms" (in our general definition, standing for atoms, ions, or molecules) are transported form the solution to form a shell around a pre-formed core, and we assume that the core and shell FCC lattice structures match, which is a good approximation for the experimentally relevant core and shell feature sizes, and have cubic lattice spacing (of FCC) $2\ell$. The pointlike atoms in solution undergo continuous-space (off-lattice) diffusion modeled as hopping at random angles, for each atom per each time steps, in steps of length for convenience set to $\sqrt{2}\ell$.

The atoms can be captured at vacant lattice sites that are nearest-neighbors of the growing structure. Each vacant site is delineated by its Wigner-Seitz unit-lattice cell, and if an atom hops into this cell, it is captured, i.e., positioned "registered" exactly at the lattice site at the center of the cell. The on-surface restructuring rules, addressed shortly, preserve this precise "registration" with the lattice structure. Note that when diffusing atoms attempt to hop into cells which are already occupied at their centers, the hopping attempt is rejected. All the attached atoms can not only move on the surface but also detach and join the diffusing atom "gas." The "registration" property of the attached atoms is crucial[26,29,30] for the emergence of the morphologies of interest. Indeed, it prevents the formation of large, structure-spanning defects that can have a "macroscopic" effect by dominating the dynamics of the shape/feature growth as a whole, e.g., by preferentially driving the growth of some crystalline faces or sustaining unequal-proportion shapes. Morphologies of interest here are obtained in the regime of growth in which such "large" defects are dynamically avoided/not nucleated, and this property is *mimicked* by the "exact registration" rule.[26,29,30]

For a typical simulation the initial shape of the FCC (gold, in experiments[1,2]) core particle on which the shell in formed (platinum, silver) was taken as an equilibrium Wulff configuration with faces formed by the (100) and (111) type crystalline planes of equal per surface atom free energy, and therefore with all the faces equidistant from the particle center, see Fig. 1 (top panel). Here the particle spans 125 FCC lattice spacings along the cartesian coordinate axes (Fig. 1), which corresponds to $250\ell$.

The growing particle is positioned in the center of a numerical-simulation region, which is a cube of size $1250\ell$. As the particle collects atoms by growing its shell, diffusional flux developed towards it in the cube. This is reminiscent of the situation encountered in applying this model to growth on a planar substrate, for which diffusional fluxes have been studied in some detail.[30] As long as the particle is small as compared to the cube, an approximately constant flux (mimicking constant concentration of diffusers in the surrounding medium) can be achieved by keeping the concentration of free (diffusing) atoms at the box's boundary, $n_0$ , fixed. The number of diffusers, $N$, in a the boundary layer of thickness 4 at the box's outer faces was calculated during the simulation and its value replenished by randomly adding additional free atoms in that layer, to keep the concentration in it equal

$$n_0 = 2.68 \times 10^{-8} N_0, \tag{1}$$

where $N_0$ thus offers a parameter to vary in order to study the effect of the solution concentration on the shell growth and morphology. The specific values used for $N_0$ and other parameters, to be defined shortly, will be addressed in the next section.

The model assumes that the atoms in the particle (core and shell) can hop to their nearest-neighbor vacant lattice sites without losing contact with the main structure. They can also detach, rejoining the diffuser population. The dynamical rules here follow those in the earlier works.[26,29,30] The hopping probabilities for each atom that is movable (not fully blocked by neighbors) mimic thermal-type transitions and are taken proportional to Boltzmann factors, which allows us to study temperature dependence of the growth. These dynamical rules are not based on the actual physical interactions, for instance those of the Ag or Pt shell atoms with each other or with the Au atoms of the core. More realistic modeling would require prohibitive numerical resources and make it impractical to study large enough systems to observe the features of interest in surface structure morphology formation. We further comment on numerical-simulation challenges in the next section.

In a MC sweep through the system, corresponding to our unit time, $t$, step, in addition to moving each freely diffusing atom we also attempt to move each lattice atom that has vacant neighbor site(s). Such an atom will have a coordination number $m_0 = 1, \dots, 11$ (for FCC). We assume that the probability for an atom to actually move during a time step is given by $p^{m_0}$, i.e., that there is a free-energy barrier, $m_0 \Delta > 0$, such that $p \propto e^{-\Delta/kT} < 1$. If the considered atom actually hops, it will end up at one of its $12 - m_0$ vacant neighbor sites with the probability proportional to $e^{m_f|\varepsilon|/kT}$, of course properly normalized over all the available target sites. Here $\varepsilon$ < 0 is the free-energy measuring binding at the prospective target site(s), the final-state coordination of which once occupied if selected, will be $m_f = 1, ..., 11$ for hopping, and $m_f = 0$ for detachment. In the latter case (detachment) the atom will proceed to freely diffuse in space after this time step.

Thus, the atom motion is determined by two parameters. Random hopping on the surface structure, with surface diffusion coefficient related to $p$, involves a (free-)energy scale $\Delta$, such that

$$p \propto e^{-\Delta/kT}. \tag{2}$$

Additional (free-)energy scale, $\varepsilon$, related to local binding, can be put in the dimensionless form by defining the parameter

$$\alpha = |\varepsilon|/kT. \tag{3}$$

Based on earlier studies we expect nonequilibrium shell formation to be well mimicked with the representative parameter values that can be set to $\alpha_0 = 1$ and $p_0 = 0.7$ to correspond to higher temperatures, and then varied with $\alpha$ up to 2.5 to decrease the temperature, with $p$ appropriately adjusted according to

$$p = (p_0)^{\alpha/\alpha_0}. \tag{4}$$

This assumes that the energy scales $\Delta$ and $|\varepsilon|$ both remain approximately constant in the considered temperature range. Note that even though we keep track of the original core atoms vs. the added shell atoms in the system, in the present modeling work we did not actually distinguish the dynamics of the core vs. shell atoms in their motion, both diffusive (some core atoms can detach) and on-surface. We also comment that the present model does not account for the temperature dependence of the diffusion constant of free atoms, which typically affects the overall transport rates and can be largely absorbed, for example, by adjusting the solution concentration.

As emphasized in earlier studies, the present kinetic MC model of particle and surface-feature growth is "cartoon" in the sense that it can only capture general qualitative features of the growth process. This limitation is shared with all the other "mesoscopic" statistical-mechanics models of this sort. The parameters that are most straightforward to vary, in order to probe their effect on the dynamics, represent "temperature," here $\alpha$ (with $p$ appropriately adjusted, see Eq. 4), and the supply of atom-size matter. The latter can be controlled by the value of $N_0$ (reflecting the "solution concentration") and of course the simulation time, $t$.

## III. Results and Discussion

Within the described modeling framework, Fig. 1 illustrates a high-temperature ($\alpha = 1$) simulation for two cases of parameter selection: lower concentration for longer deposition time vs. five-fold higher concentration but five-fold shorter process. Figure 1 illustrates one key finding of our modeling results. With proper selection of the parameters, we can obtain relatively smooth epitaxial shells. This observation is experimentally very recent[1,2,24,25] as compared to earlier experimental works, and it is important for new functionalities in applications of core-shell particles. The shell and core atom intermixing (only shell atoms are shown in the middle panel of Fig. 1) was found to be present but not significant. The diffusional intake of matter is larger in regions of higher curvature, and therefore it is not surprising that the shell gradually distorts from the original core by bulging out at the corners and edges of the original shape, as seen in the bottom panel.

However, recall that we are working in the nonequilibrium growth regime of comparable time scales of adding matter vs. its on-surface redistribution by dynamical processes described in the preceding section. The flow of matter on-surface, away from the bulging regions suffices in some situations to keep the shell from too rapidly destabilizing. The shell then remains smooth. This is observed for larger temperatures in our nomenclature, likely because the on-surface diffusion is faster. Indeed, the range for the temperature variation considered here, with, as mentioned, $\alpha$ from 1 up to 2.5, is set based on earlier work[26,28-30] with this model, and Fig. 1 corresponds to the largest temperature, $\alpha = 1$.

The distortion of the particle shape is increased if we attempt to increase the particle concentration, by taking larger $n_0$ (increased by increasing $N_0$, see Eq. 1) to shorten the process time. This is seen in the middle panel of Fig. 1. Thus, smoother shells will be obtained in relatively higher-temperature conditions, and in more dilute solutions, the latter, however, requiring longer synthesis times and potentially resulting in somewhat more intermixing of the shell and core materials (observed in the middle panel). The need to slow down the atom supply rate to get smooth epitaxial shell was noted in experiments.[1,2]

Figure 2 offers an illustration of the onset of another shell-formation regime. Here the temperature was lowered as compared to Fig. 1, but the other parameters are similarly selected. In this case (Fig. 2) instabilities in growth appear on the time scales for which the shell was smooth for larger temperature. In the case of lower $n_0$ (but larger deposition time, the top panel in Fig. 2), clustering/clumping occurs primarily on the (111) type FCC faces of the core. However, as $n_0$ was increased (and the process was ended faster, the bottom panel), the unstable growth was seen both on the (111) and (100) type FCC faces, and the characteristic dimensions of the clusters/clumps were smaller.

This type of cluster-structured shell is observed in many experiments.[1-12] However, in real experiments small clusters can also nucleate in solution and then aggregate on the growing particles. It is likely that both mechanisms, local clustering and also aggregation of separately nucleated clusters, contribute to the final shell morphologies. We will comment later in this section on the extent to which the local clustering mechanism such as that seen in Fig. 2 can on its own produce morphological features similar to those experimentally observed. Our model does not include off-particle cluster nucleation: We consider only a single "effective" particle with supply of atoms driving its growth. This is done because small clusters' aggregation can typically play role in the formation of non-epitaxial shells only. Modeling of the combined processes would require a multi-scale approach and is outside the scope of any presently numerically tractable kinetic MC setting. Our primary aim here has been to develop modeling particularly suited for the smooth epitaxial shell formation, though we also describe the morphology of the shells with cluster structure that are formed by the same processes in different kinetic regimes.

To highlight the numerical challenges, we report that total amount of computational effort for the present project is equivalent to approximately half a year of CPU time on modern multi-core clusters with simultaneously running 30–40 cores, requiring 120–160 GB of dedicated memory. We utilized low-level parallelization for multi-threading, similar to OpenMP, and we actually used several clusters with CPUs such as Intel® Xeon® E5420 (2.5 GHz), Intel® Core™ i7-870 (2.93 GHz), etc. The longest numerical runs were for low $\alpha$ ($= 1.0$), with a large supply of atoms, $N_0 = 2 \times 10^5$, and $2 \times 10^7$ to $3 \times 10^7$ atoms being added to the system (the initial particle contains approximately $5.3 \times 10^6$ atoms). On a single-core processor it would take approximately half a year of CPU and 5 GB of memory just to simulate the first $10^6$ MC steps (sweeps through the system) in such cases.

In order to elucidate the onset of clustering driven by the supply of atoms, let us consider the initial stages of the shell formation. Figure 3 shows illustrative examples of low-temperature (top panel) vs. high-temperature (bottom panel) growth for relatively thin shells. We note that a thin shell can be rather smooth if the flux of matter is not too large and the temperature is high enough for fast on-surface matter redistribution. However, when the temperature is lowered, instabilities begin to form as clusters, reminiscent of the behavior observed for growth on planar substrates.[29,30] Interestingly, the clusters form even before the core is completely obscured by the shell. This property correlates with the experimental observation[2] that electrochemical activity of the core metal (gold) was observable in the cluster-shell case, but not for the smooth-shell particles, in addition to the electrochemical signature of the shell metal (platinum). This of course does not preclude the additional aggregation of solution-nucleated small clusters in the final cluster-structured shell, but our observation indicates that the atom-capture mechanism alone can qualitatively reproduce the observable shell morphology properties, at least for thin cluster-structured shells.

Consideration of larger-time (thicker) shells formed in the cluster-regime, see Fig. 4, indicates that the configuration shown in the top panel of Fig. 3 can provide the substrate for the emergence of a thick cluster-structured shell (the top panel in Fig. 4). For a larger solution density (and respectively shorter process time) the clustering scales can be made smaller (the bottom panel in Fig. 4), which presumably reflects a stronger tendency for clustering. In fact, these larger-time configurations, especially the one shown in the top panel (Fig. 4) are visually very similar to the experimental snapshots of shell morphologies, cf. Fig. 4 in Ref. 2. Further similarity of the onset of clustering here as compared to the growth on planar substastres[29,30] is highlighted in Fig. 5, where the morphology of small-scale clustering-initiating instabilities that grow as pyramid-shaped islands is shown, which should be compared to similar results results for planar substrates, reported in Ref. 29.

In summary, we applied a kinetic MC model earlier used for other related particle/cluster growth studies to the shell formation in core-shell particle synthesis. We observed that several key features of the recently experimentally explored systems can be qualitatively reproduced. These include the formation of smooth epitaxial shells in some regimes, especially for higher-temperature but lower-solution concentration deposition processes. Formation of cluster-structured shells was also observed, including the property that such shells can in some growth regimes leave sizable parts of the core exposed, and that morphologies visually similar to those experimentally reported are reproducible.

We thank Prof. D. V. Goia and Dr. I. Sevonkaev for useful discussions.

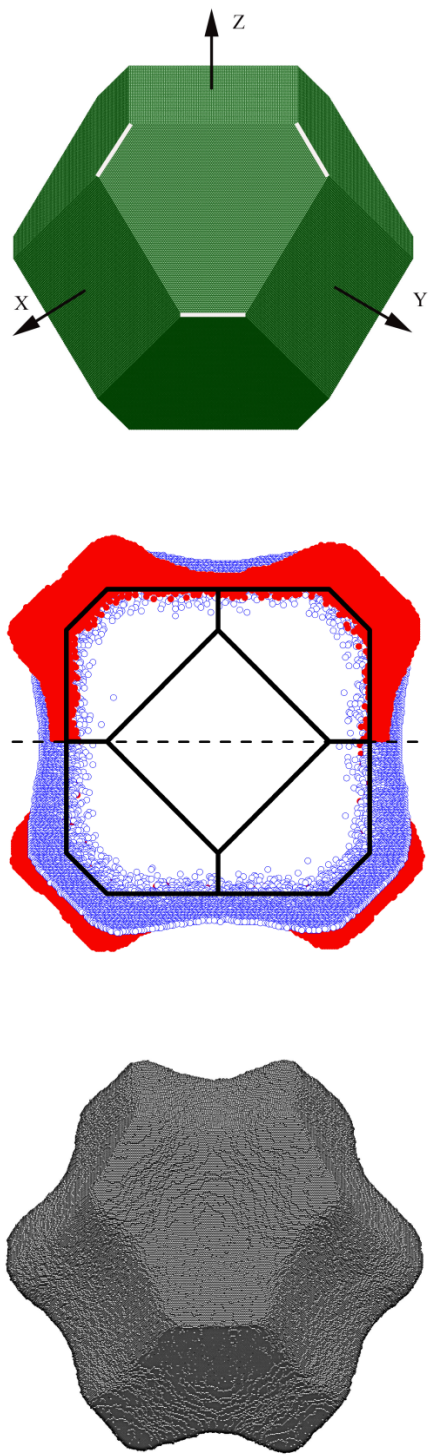

**Figure 1.** *Top panel:* The initial shape of the core nanoparticle that spans 125 FCC lattice spacings along the shown coordinate axes (see text for details). The white segments where added

to highlight some of the regions of the highest mean curvature that constitute the shape's edges. *Middle panel:* The distribution of the deposited atoms in the shell as seen by depicting the $z = 0$ plane cross-section, shown for two different numerical simulation runs, color coded as follows. Blue atoms correspond to the case $\alpha = 1$, $N_0 = 4 \times 10^4$, $t = 5 \times 10^6$, red atoms correspond to the case $\alpha = 1$, $N_0 = 2 \times 10^5$, $t = 10^6$, where the model parameter values are defined in the text. Solid lines show the (projections of the) edges of the original core. (Note that the images in the panels here and in other figures are all at somewhat different magnifications to have approximately fixed horizontal sizes. They should not be compared to each other's sizes.) In both simulations shown in the middle panel the total number of the deposited shell atoms was about $3.5 \times 10^6$, whereas the core initially had approximately $5.3 \times 10^6$ atoms. The shell (only) atoms are depicted, as small circles, with different-color shell atoms superimposed on top of the other color in the upper and lower halves of the middle-panel image. *Bottom panel:* The overall view of the core-shell structure at the end of the simulation, for the case $N_0 = 2 \times 10^5$.

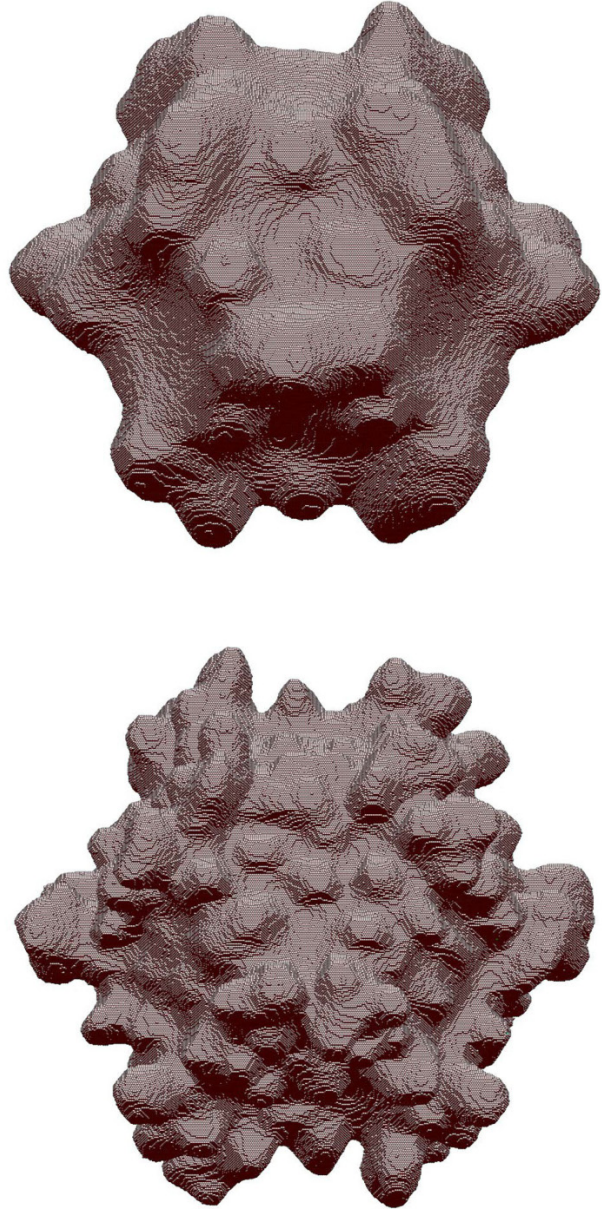

**Figure 2.** *Top panel:* Particle shape obtained for growth with parameter values $\alpha = 2$, $N_0 = 4 \times 10^4$, $t = 5 \times 10^6$. The core was initially the same as in Fig. 1. *Bottom panel:* Shape obtained for the same value of $\alpha = 2$, but with $N_0 = 2 \times 10^5$, $t = 10^6$. In both cases approximately $4.4 \times 10^6$ atoms were deposited.

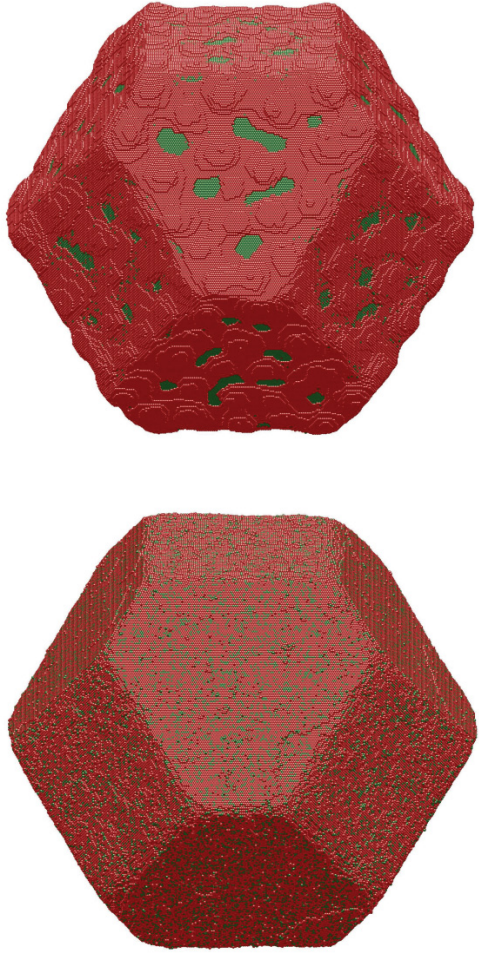

**Figure 3.**      *Top panel:* Onset of clustering in growth at low temperatures, for $\alpha = 2.5$, $N_0 = 2 \times 10^4$, $t = 10^6$. Here the core remains practically intact. The atoms originally in the core are color coded in green, whereas the added atoms are color coded in burgundy. The shell is primarily formed by the added atoms and its growth morphology is initially governed by the emergence of ziggurat-type clusters. For the selected process time the shell contains approximately 6% of atoms when compared to the number of atoms in the core. *Bottom panel:* Absence of any significant clustering at high temperatures. Here $\alpha = 1$, with the same $N_0 = 2 \times 10^4$, but the simulation time was taken as $t = 5 \times 10^6$. The growing surface remains rather smooth even for the large time selected, and it includes a significant admixture of atoms from the original core. For this process time the shell here contains about 25% atoms when compared to the initial count of atoms in the core.

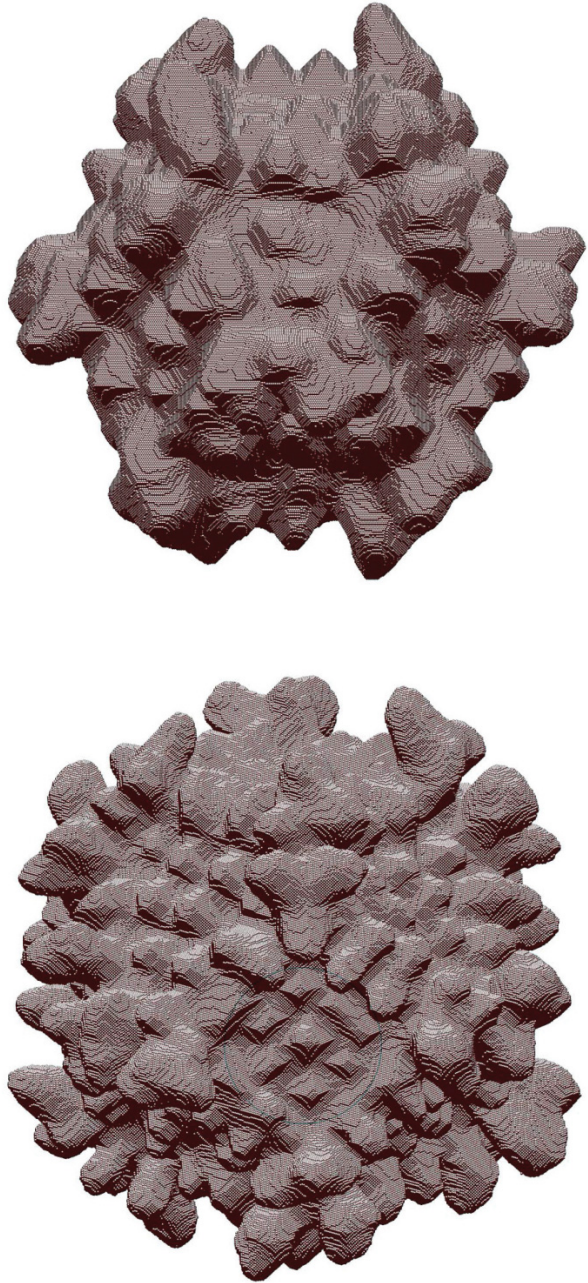

**Figure 4.** *Top panel:* Development of clustering for the same particle as shown in the top panel of Fig. 3, but for large time. Here $\alpha = 2.5$, $N_0 = 2 \times 10^4$, $t = 8 \times 10^6$. *Bottom panel:* Smaller-scale clustering for the case of $\alpha = 2.5$, $N_0 = 6 \times 10^4$, $t = 3 \times 10^6$.

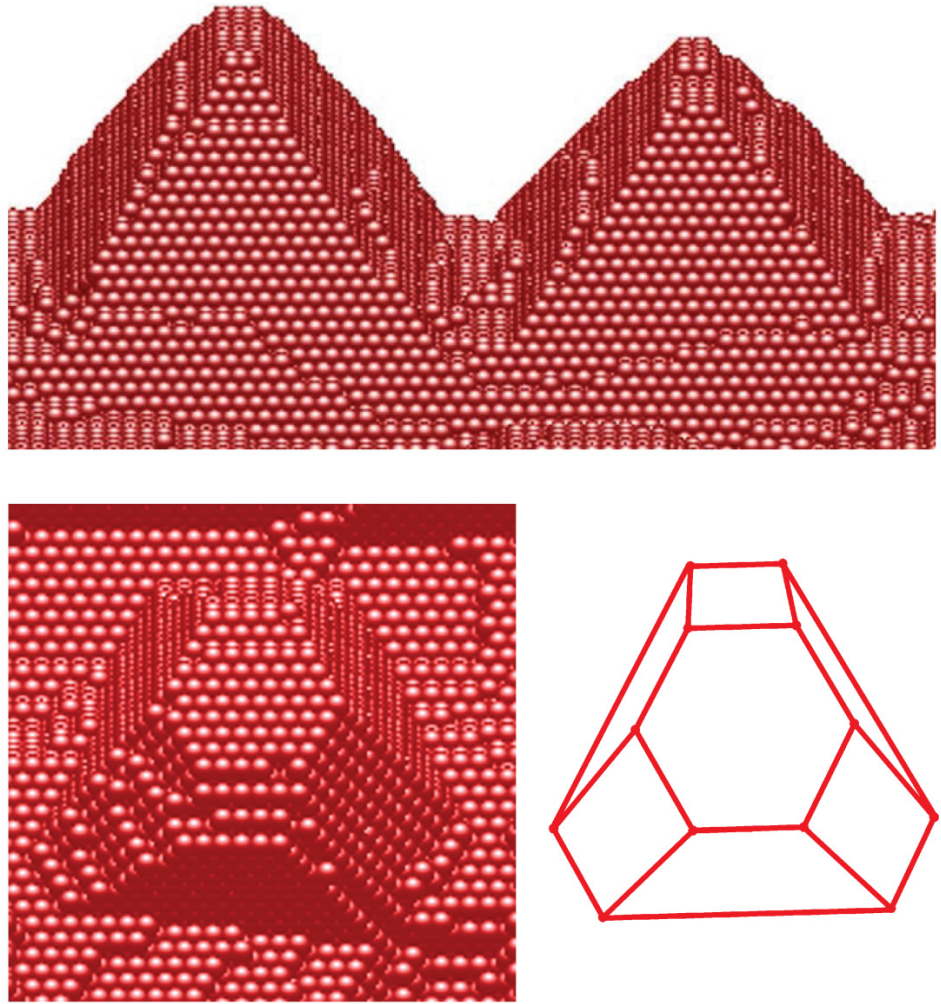

**Figure 5.** Illustration of the morphology of pyramid-shaped instabilities initiating the irregular cluster-mode growth in situations such as those presented in Fig. 4, and also in the top panel in Fig. 3 (for larger times than shown there). *Top panel:* Pyramidal clusters developing on growing (100) type FCC faces, with all the side faces of the type (111). *Bottom panel:* Pyramidal clusters developing on growing (111) type FCC faces, compared to a schematic depicting a pyramidal shape made of fragments of various (111) and (100) FCC faces.

**Table of Contents Image**

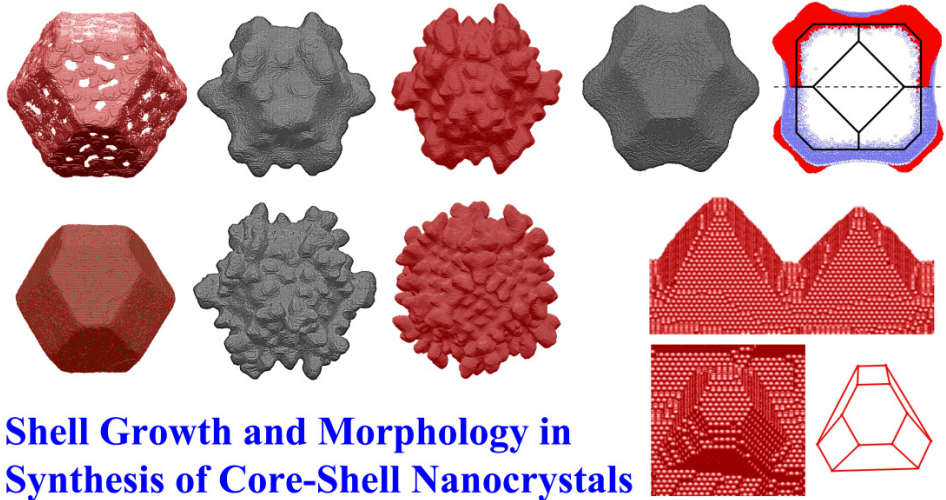

**Shell Growth and Morphology in Synthesis of Core-Shell Nanocrystals**

# Journal Issue Cover

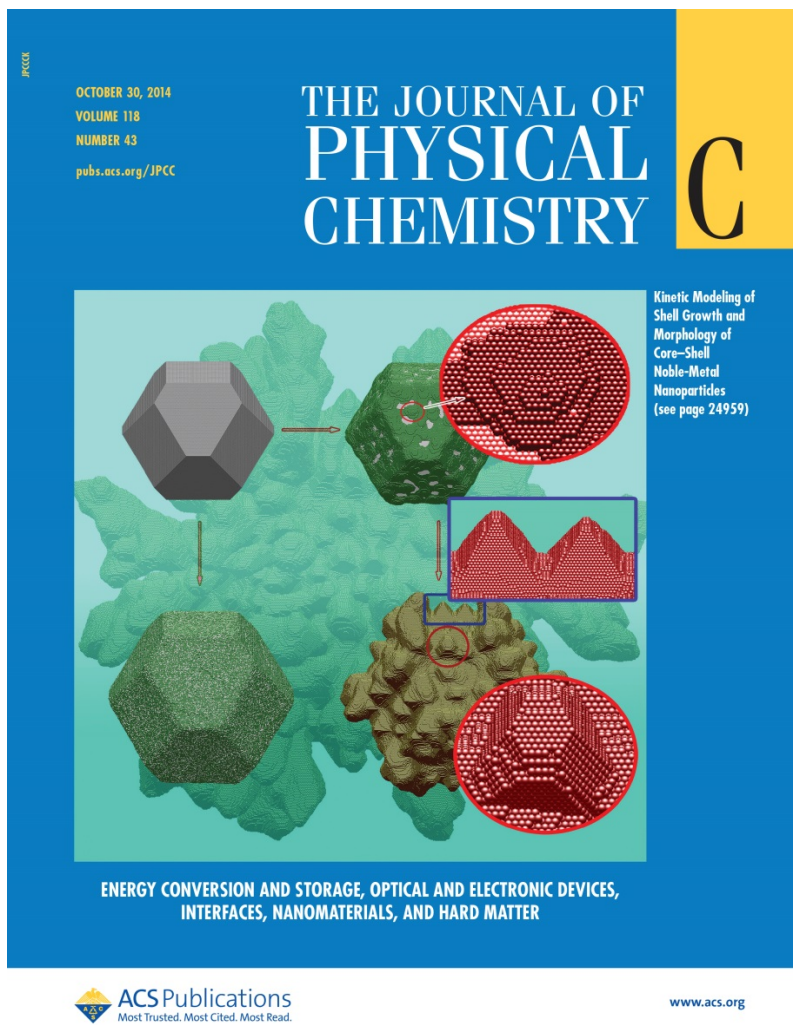

**Legend:** Kinetic modeling of shell growth and morphology of core–shell noble-metal nanoparticles. The kinetic Monte Carlo modeling approach is shown to reproduce smooth-shell and cluster-structured shell growth morphologies identified in recent experiments on core–shell noble-metal nanoparticle synthesis, including the formation of smooth epitaxially grown shells. The effects of temperature and supply of matter on the resulting shell morphology are considered for growth on presynthesized nanocrystal cores.